\documentclass{PoS}
\def\inlinetilde{\lower0.8ex\hbox{$\,\widetilde{}\,$}}
\def\chpt{\raise0.4ex\hbox{$\chi$}PT}
\def\schpt{S\raise0.4ex\hbox{$\chi$}PT}
\def\rschpt{rS\raise0.4ex\hbox{$\chi$}PT}
\def\figref#1{Fig.~\ref{fig:#1}}
\def\Figref#1{Figure~\ref{fig:#1}}

\def\tabref#1{Table~\ref{tab:#1}}

\def\gtwid{{\,\raise.3ex\hbox{$>$\kern-.75em\lower1ex\hbox{$\sim$}}\,}}
\def\ltwid{{\,\raise.3ex\hbox{$<$\kern-.75em\lower1ex\hbox{$\sim$}}\,}}

\def\circle{{\Large{\raise-0.15ex\hbox{$\circ$}}}}
\def\sqar{{\raise-0.1ex\hbox{$\Box$}}}   %scales with fonts
\def\ie{{\it i.e.},\ }
\def\eg{{\it e.g.},\ }
\def\et{{\it et al.}}

\def\eqn#1{\label{eq:#1}}

\def\nsection#1 #2{\leftline{\rlap{#1}\indent\relax #2}}

\def\prd#1{Phys.\ Rev.\ D {\bf #1}}

\def\fermilabtwo{Nucl.\ Phys.\ {\bf B} (Proc.\ Suppl.) {\bf 140} (2005)}

\def\dublin#1{PoS {\bf LAT2005}, #1,\ (2005) {\tt http://pos.sissa.it}}
\def\tucson#1{PoS {\bf LAT2006}, #1,\ (2006) {\tt http://pos.sissa.it}}

\def\regensburg#1{presented at the International Symposium,
{\it Lattice 2007}, Regensburg, Germany, July 30--Aug.~4, 2007, 
PoS {\bf LAT2007}, #1}
\def\MeV{{\rm Me\!V}}
\def\GeV{{\rm Ge\!V}}
\def\msbar{{\overline{\rm MS}}}

\def\dublin#1{PoS {\bf LAT2005}, #1,\ (2005)}

\def\tucson#1{PoS {\bf LAT2006}, #1,\ (2006)}

\title{Status of the MILC light pseudoscalar meson project}

\ShortTitle{MILC light pseudoscalar project}

\author{\speaker{C.\ Bernard}\\
        Washington University; Saint Louis, Missouri, USA\\
        E-mail: \email{cb@lump.wustl.edu}}

\author{C.\ DeTar and L.\ Levkova\\
	University of Utah; Salt Lake City, Utah, USA\\
	E-mail: \email{detar@nova.physics.utah.edu},
		\email{ludmila@phys.columbia.edu}}

\author{Steven Gottlieb\\
	Indiana University; Bloomington, Indiana, USA\\
	E-mail: \email{sg@indiana.edu}}

\author{U.M.\ Heller\\
	American Physical Society; Ridge, New York, USA\\
	E-mail: \email{heller@csit.fsu.edu}}

\author{J.E.\ Hetrick\\
 University of the Pacific; Stockton, California USA\\
	E-mail: \email{jhetrick@pacific.edu}}

\author{J.\ Osborn\\
 Boston University; Boston, Massachusetts USA\\
	E-mail: \email{josborn@physics.bu.edu}}
 
\author{D.\ Renner and D.\ Toussaint\\
University of Arizona; Tucson, Arizona, USA\\
	E-mail: \email{dru@physics.arizona.edu},
		\email{doug@klingon.physics.arizona.edu}}

\author{R.\ Sugar\\ 
University of California; Santa Barbara, California, USA\\
	E-mail: \email{sugar@physics.ucsb.edu}}

\abstract{
We discuss the current status of our calculation of the physics
of $\pi$ and $K$ mesons using three dynamical flavors of improved staggered
quarks.  This year, we have a new ensemble with a lattice spacing 
of $0.06\;$fm and a light sea mass of $0.2 m_s$, as well as significant 
increases in statistics at several coarser lattice
spacings and/or heavier sea masses. Results for decay constants, 
quark masses, low energy constants, condensates, and $V_{us}$ are presented.
}

\FullConference{The XXV International Symposium on Lattice Field Theory\\
		 July 30-4 August 2007\\
		 Regensburg, Germany}

\begin{document}

We are using improved staggered quarks \cite{ASQTAD} with $N_f=3$ dynamical flavors (both 
unquenched (``full'') QCD and 
partially quenched) to study the physics of light pseudoscalars ($\pi$, $K$).
Since our original published work \cite{FPI04}, we have continued to add data sets
with lighter sea quark masses and/or finer lattice spacings and to improve the
analysis.  This is the latest in a series of 
periodic updates \cite{FPILAT04,FPILAT05}.
We concentrate here on those aspects that have changed since last year.

%A study  of the $\pi$-$K$ system makes possible:
%
%\begin{itemize}
%\item[{$\bullet$}\ ]{}
%A precise extraction of the CKM element $V_{us}$ from
%$f_K/f_\pi$, competitive with the world-average.
%
%\item[{$\bullet$}\ ]{}
%A determination of the light quark masses and their
%ratios with high lattice precision.
%
%\item[{$\bullet$}\ ]{}
%A test of the applicability of rooted staggered chiral perturbation theory (\rschpt) \cite{LEE-SHARPE,SCHPT-CACB,SCHPT-OTHER}
%for describing staggered lattice data.
%
%\item[{$\bullet$}\ ]{}
%A sensitive check of our algorithms and methods --- including
%the $\root 4 \of {\rm Det}$ trick for dynamical staggered quarks --- 
%by comparing $f_\pi$ to the well-determined experimental value.
%
%\item[{$\bullet$}\ ]{}
%A determination  of low energy constants (LECs), such as the Gasser-Leutwyler
% $L_i$ and condensates such $\langle \bar\psi\psi\rangle$ in the two- and three-flavor chiral limits.
%
%\end{itemize}
%

\tabref{lattices} gives the parameters of our lattices.
The quantities $m'_s$ and $\hat m' \!=\! m'_u\!=\!m'_d$ denote the
values of sea quark masses chosen in each run.  
(The corresponding masses without the primes, \eg $m_s$ and $\hat m \equiv (m_u+m_d)/2$,
are the physical values.)
    
\begin{table}[t!]
%\begin{table}[ptbh!]
%\begin{table}[b]
\newcommand{\rs}{\phantom{R}}
\begin{center}
\setlength{\tabcolsep}{3mm}
\begin{tabular}{|c|c|c|c|c|c|c|c|}
\hline
$a$ (fm) & $a\hat m'$ / $am'_s$ \rule[0mm]{0mm}{4mm} & $L$ (fm) & $m_\pi$ / $m_\rho$ & $m_\pi L$ & $10/g^2$ & Lat Dim  & \# Lats \\
\hline                            
%\noalign{\vspace{1.5mm}}
%\multicolumn{6}{c}{$a\approx 0.15$ fm (``coarser'')\rule[0mm]{0mm}{4mm}}\\
\hline
%{0.0484 / 0.0484  } &{2.4 } &{0.611 } &{8.6 } &{6.628 } &{$16^3\times 48$ } &{600 }\\
$\approx\!0.15$ & {0.0290 / 0.0484 } &{2.4 }  &{0.522 } &{6.7 } &{6.600 } &{$16^3\times 48$ } &{600 }\\
$\approx\!0.15$ & {0.0194 / 0.0484 }  &{2.4 } &{0.454 } &{5.5 } &{6.586 } &{$16^3\times 48$ } &{600 }\\
$\approx\!0.15$ & {0.0097 / 0.0484 }  &{2.4 } &{0.348 } &{3.9  } &{6.572 } &{$16^3\times 48$ } &{600 }\\
$\approx\!0.15$ & {0.00484 / 0.0484 }  &{3.0 } &{0.256 } &{3.4  } &{6.566 } &{$20^3\times 48$ } &{600 }\\ 
\hline                            
%\noalign{\vspace{1.5mm}}
%\noalign{\vspace{4mm}}
%\multicolumn{6}{c}{$a\approx 0.12$ fm (``coarse'')\rule[0mm]{0mm}{4mm}}\\
\hline
%0.40/0.40   & 0.94 & 29.4  & 7.35 & $20^3\times 64$ & 332   \\
%0.20/0.20   & 0.89 & 19.6  & 7.15 & $20^3\times 64$ & 341   \\
%0.10/0.10   & 0.79 & 13.7  & 6.96 & $20^3\times 64$ & 339  \\
%0.05/0.05   & 0.68 & 9.7  &6.85 &  $20^3\times 64$ & 425   \\
%0.04/0.05   & 0.63 & 8.7  & 6.83 & $20^3\times 64$ & 351   \\
$\approx\!0.12$ & {0.03  / 0.05  }  &{2.4 } & {0.582  } & {7.6 } & {6.81  } & {$20^3\times 64$  } & {362 }\\
$\approx\!0.12$ & {0.02  / 0.05   }  &{2.4 } & {0.509 } & {6.2  } & {6.79 } & {$20^3\times 64$  } & {485 }\\
$\approx\!0.12$ & {0.01  / 0.05   }  &{2.4 } & {0.394 } & {4.5  } & {6.76 } & {$20^3\times 64$ } & {894 }\\   
$\approx\!0.12$ & {0.01  / 0.05   }  &{3.4 } & {0.395 } & {6.3  } & {6.76 } & {$28^3\times 64$  } & {275 }\\   
$\approx\!0.12$ & {0.007  / 0.05  }  &{2.4 } & {0.342 } & {3.8  } & {6.76 } & {$20^3\times 64$ } & {836 }\\   
$\approx\!0.12$ & {0.005  / 0.05  }  &{2.9 } & {0.299  } & {3.8  } & {6.76 } & {$24^3\times 64$ } & {527 }\\
$\approx\!0.12$ & {0.03  / 0.03   }  &{2.4 } & {0.590   } & {7.6   } & {6.81  } & {$20^3\times 64$  } & {360 }\\
$\approx\!0.12$ & {0.01  / 0.03    }  &{2.4 } & {0.398  } & {4.5    } & {6.76  } & {$20^3\times 64$  } & {349 }\\ 
\hline
%\noalign{\vspace{4mm}}
%\noalign{\vspace{1.5mm}}
%\multicolumn{6}{c}{$a\approx 0.09$ fm (``fine'')\rule[0mm]{0mm}{4mm}}\\
\hline
$\approx\!0.09$ & {0.0124  / 0.031 }  &{2.4 } & {0.495 } & {5.8  } & {7.11  } & {$28^3\times 96$  } & {531 }\\   
$\approx\!0.09$ & {0.0062  / 0.031 }  &{2.4 } & {0.380 } & {4.1 } & {7.09  } & {$28^3\times 96$ } & {583 }\\   
$\approx\!0.09$ & {0.0031  / 0.031 }  &{3.4 } & {0.297 } & {4.2 } & {7.08  } & {$40^3\times 96$ } & {503 }\\   
\hline
%\noalign{\vspace{4mm}}
%\noalign{\vspace{1.5mm}}
%\multicolumn{6}{c}{$a\approx 0.06$ fm (``super fine'')\rule[0mm]{0mm}{4mm}}\\
\hline
$\approx\!0.06$ & {0.0072 / 0.018 }  &{2.9 } & {0.474 } & {6.3 } & {7.48 } & {$48^3\times 144$ } & {556 } \\
$\approx\!0.06$ & {0.0036 / 0.018 }  &{2.9 } & {0.370 } & {4.5 } & {7.47 } & {$48^3\times 144$ } & {334 }\\ 
\hline
\end{tabular}
\end{center}
\vspace{-4mm}
\caption{\label{tab:lattices}
Lattice parameters.  The lattice spacings are the ``nominal'' scales (see text).
%The $a\approx 0.15\;$fm lattices are known as the ``coarser'' sets;
%$a\approx 0.12\;$fm, ``coarse'';
%$a\approx 0.09\;$fm, ``fine''; and
%$a\approx 0.06\;$fm, ``super fine''.
The $\pi$ and $\rho$ 
referred to are those formed out of
the sea quarks for each lattice; valence quark masses however go 
down to the lightest sea-quark values in 
the table.}
\vspace{-4mm}
\end{table}

The $a\!\approx\! 0.06\;$fm lattice with masses
{0.0036 / 0.018} is a new ensemble this year, as is (for this analysis) the large-volume
$a\!\approx\! 0.12\;$fm lattice with masses
{0.01 / 0.05} and spatial size {$28^3$}.  
The numbers of configurations for the $a\!\approx\! 0.06\;$fm lattice with masses
{0.0072 / 0.018} and for  several of the $a\!\approx\! 0.12\;$fm lattices
have almost doubled since last year.  
Running on an  $a\!\approx\! 0.06\;$fm lattice
with $\hat m' = 0.1 m_s'$ (masses 0.0018 / 0.018) has recently begun but is not
included here.

On each ensemble, we determine $r_1/a$, where $r_1(\hat m', m_s', g^2)$  \cite{MILC-POTENTIAL} is 
a length scale from the static quark potential, similar to $r_0$ \cite{SOMMER}.
The quantity  $r_1^{\rm phys}$, defined as the continuum $r_1$ 
at physical quark masses ($\hat m$, $m_s$) may be determined
from the $r_1/a$ values and the $\Upsilon$ 2S-1S splitting \cite{HPQCD-UPSILON}.
We obtain $r_1^{\rm phys}=0.318(7)\;$fm \cite{FPILAT05,MILC-SPECTRUM}. 

For generic chiral and continuum extrapolations, it is convenient to
define the lattice scale by 
$a\equiv r_1^{\rm phys}/(r_1(\hat m',m_s', g^2)/a)$.
We call this the ``nominal'' scale-setting procedure.
In choosing the input lattice coupling $g^2$, we kept $r_1/a$ fixed as 
$\hat m'$ and $m_s'$ changed over a given set of ensembles (\eg the $a\!\approx\! 0.12\;$fm ensembles).
Thus, up to tuning errors, each ensemble grouped within a box in
\tabref{lattices} has the same nominal scale.
However, fixing the scale this way is not completely correct for applying chiral
perturbation theory (\chpt) to quantities such as $f_\pi$
since $r_1$ has some (small, but physical) dependence on the dynamical
quark masses that is not included in \chpt. 

A mass-independent 
procedure to set the scale is preferable \cite{Sommer:2003ne}.
A convenient procedure is
to replace $r_1(\hat m',m_s', g^2)/a$ by $r_1(\hat m,m_s, g^2)/a$, where
the value of $r_1(\hat m,m_s, g^2)/a$ at physical masses $\hat m,m_s$ 
is obtained by a smooth interpolation/extrapolation from $r_1(\hat m',m_s', g^2)/a$.
We tried this mass-independent scheme in Ref.~\cite{FPI04},  but the differences with
the nominal approach were smaller than other systematic errors for all quantities.
With better data, we now find significant differences in a few low energy
constants (LECs).  In addition, 
the mass-independent scheme tends to have better confidence levels in our
\chpt\ fits. Therefore we use this scheme exclusively here.

As in Refs.~\cite{FPI04,FPILAT04,FPILAT05} we fit the partially quenched (PQ) lattice
data to rooted staggered chiral perturbation theory (\rschpt) forms \cite{LEE-SHARPE,SCHPT-CACB,SCHPT-OTHER}.  We always fit multiple lattice
spacings, and both masses and decay constants, simultaneously.
To determine the LO and NLO LECs and chiral-limit quantities,
we fit 
to the low quark-mass region, and omit the $a\!\approx\! 0.15\;$fm lattices,
where taste violations are large.   
Denoting the valence quark masses in the mesons by $m_x$ and $m_y$, the low-mass cuts are:
		%$am_x+am_y \le 0.018\approx 0.39 am_s$ ({coarse})\\
                %$am_x+am_y \le 0.015 \approx 0.51 am_s$ ({fine})\\
                %$am_x+am_y \le 0.011 \approx 0.56 am_s$ ({super fine})
		$am_x+am_y \ltwid 0.39\; am_s$ (at $a\!\approx\! 0.12\;$fm);
                $am_x+am_y \ltwid 0.51\; am_s$ (at $a\!\approx\! 0.09\;$fm); and
                $am_x+am_y \ltwid 0.56\; am_s$ (at $a\!\approx\! 0.06\;$fm).
We can tolerate a higher cutoff at smaller lattice spacing because the taste violations,
and hence the masses of non-Goldstone pions, are smaller.
In these fits, we also cut on sea-quark mass and remove the $a\!\approx\! 0.12\;$fm sets with masses
$0.03/0.05$, $0.02/0.05$, and $0.03/0.03$.
                 %$m_s' + 2\hat m'\ltwid 1.5 m_s$ ({coarse})\\
                  %$m_s' + 2\hat m'\ltwid 1.9 m_s$ ({fine})\\
                 %$m_s' + 2\hat m'\ltwid 1.7 m_s$ ({super fine})
                  %$m_s' + 2\hat m'\ltwid 1.8 m_s$ always
Because the statistical errors are so small,
we still need to add in the NNLO analytic terms to the complete
NLO forms in order to get good fits \cite{FPI04}. 
% (0.1\% to 0.4\% for meson masses \& decay constants), 

%		Joint fits to decay constants and masses
%		have 26 or 27 free parameters. 
%	
%		An effective NNLO parameter is sometimes added: coefficient of one-loop chiral logs is allowed 
%to float to take into account differences
%between $f$, $f_\pi^{\rm phys}$, and $f_K^{\rm phys}$.  Best fit has coefficient $0.88/(4 \pi f_\pi^{\rm phys})^2$.

For interpolation around $m_s$, we must include higher quark masses.
Once LO and NLO parameters are determined, we fix them (up to statistical errors)
and fit to all sea mass sets, all lattice spacings, and valence masses $m_x+m_y \ltwid 1.2\; m_s$.
We now also need to  add in NNNLO analytic terms to get good fits.
These NNNLO fits are used for central values of $f_\pi$, $f_K$ and quark masses.

		%\item[{$\bullet$\ ]{} Valence mass ranges are:
		%$am_x+am_y \le 0.068\approx 1.05 am_s$ ({coarser})\\
		%$am_x+am_y \le 0.055\approx   1.19 am_s$ (coarse})\\
                %$am_x+am_y \le 0.0353 \approx 1.20 am_s$ ({fine})\\
                %$am_x+am_y \le 0.0253 \approx 1.30 am_s$ ({super fine}) 
%Total for these fits is 54 to 55 params, with 12 or 13 (LO, LO $a$ dependence, 
%NLO, and [where applicable] coefficient of chiral log) tightly
			%constrained from low-mass fits.

\begin{figure}[ht]
\begin{center}
\includegraphics[width=0.5\textwidth]{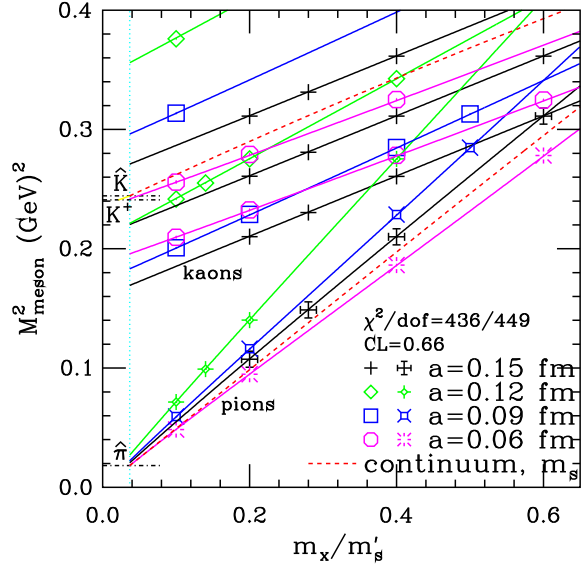}
\end{center}
\vspace{-0.6truecm}
\caption{Comparison of NNNLO fit to partially-quenched squared meson masses.  
For clarity only the lightest sea-quark ensemble for each
lattice spacing is shown.  \label{fig:masses}}
\end{figure}

\Figref{masses} shows results for the squared pseudoscalar masses
as a function of quark mass.  
``Pions'' have valence masses $m_y=m_x$;
while ``kaons'' have  $m_y$  held fixed
at various (arbitrary) values while $m_x$ varies.
The fit is to the full quark-mass range and uses NNNLO terms.

For the pions,
the relative values of the results on various lattices is determined largely by 
the relation between the simulation strange mass $m_s'$ in the sea
and the physical mass $m_s$.
For example, $m_s'/m_s$ is largest for the $a\!\approx\! 0.12\;$fm lattices, which makes the slope
of the pion data greatest for these lattices.  
For kaons, the biggest effect is simply the choice
of the values of the fixed valence mass $m_y$, typically chosen to be various fixed fractions
of $m_s'$.

Extrapolating to the continuum and setting valence and sea quark
    masses equal, we get the
dashed red lines; $m_s'$ has been adjusted so that both the kaon and the pion hit their
physical values at the same value of $m_x$.  
This gives the physical quark masses
$\hat m$ and $m_s$ (after renormalization). 

\begin{figure}
\begin{center}
\parbox[t]{0.47\textwidth}{
\includegraphics[width=0.47\textwidth]{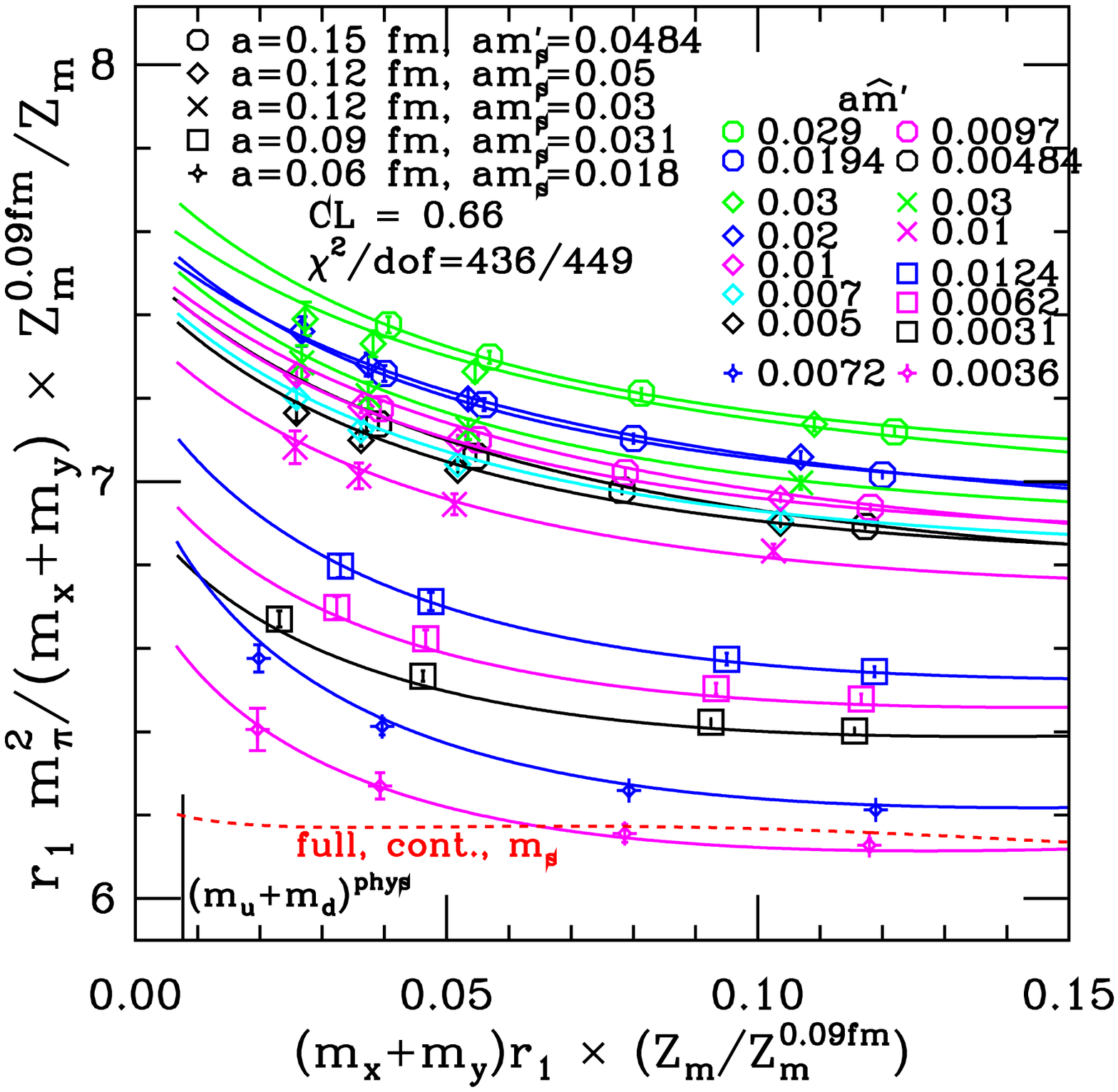}
}%
\hspace{0.2cm}
\parbox[t]{0.50\textwidth}{
\includegraphics[width=0.50\textwidth]{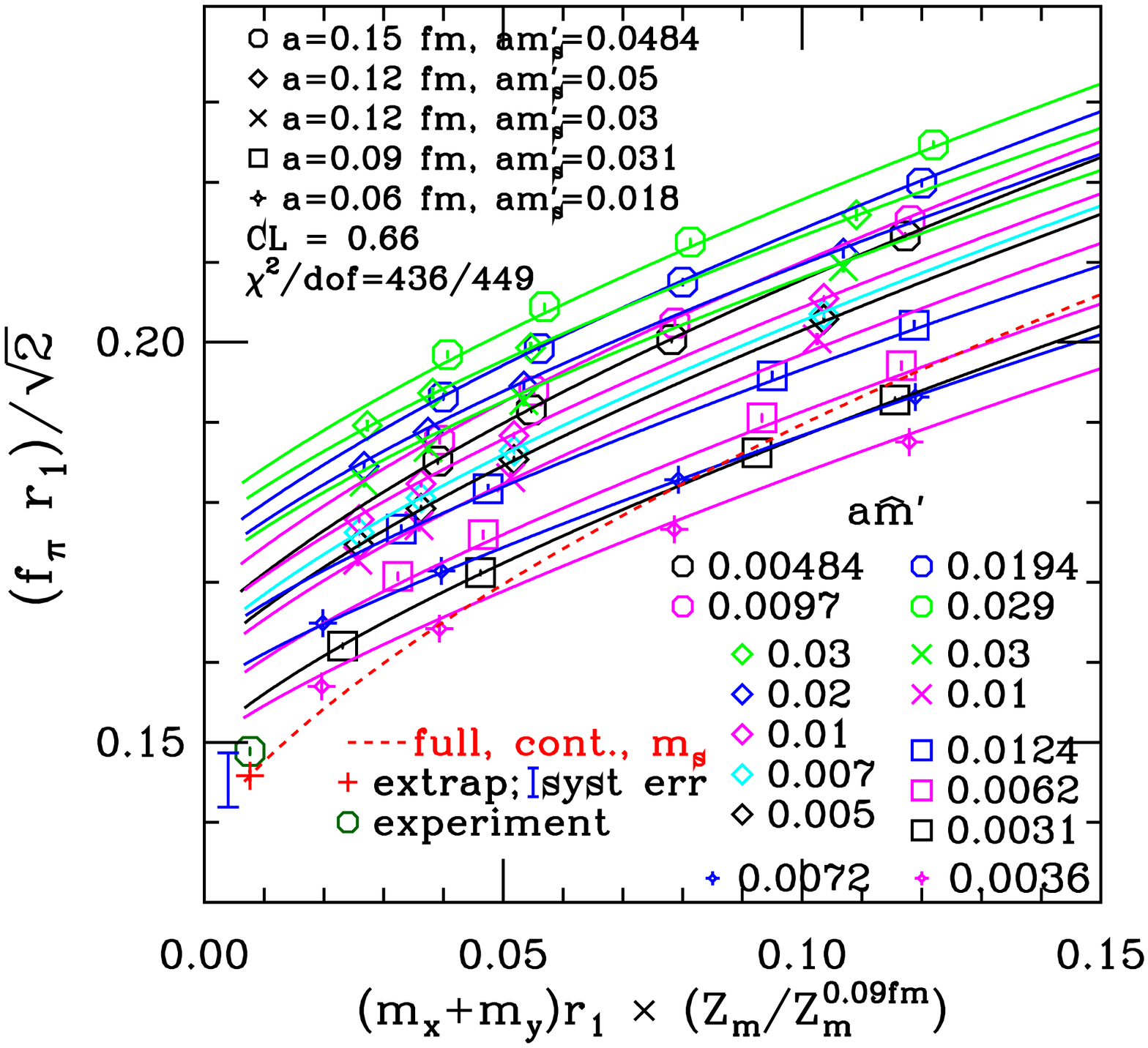}
}%
\end{center}
\vspace{-0.6truecm}
\caption{Data for
$m_\pi^2/(m_x+m_y)$ is plotted at left; while that for $f_\pi$ is plotted at right.
This is the same fit as in Fig.~1.  All sea-quark ensembles are represented, but
only ``pion'' points ($m_x\!=\!m_y$) are shown.
\label{fig:m-and-f}}
\vspace{-0.3truecm}
\end{figure}

Note that the fit
lines in \figref{masses} are remarkably straight on this scale.
To see curvature coming from 
the NLO chiral logs as well as the analytic
higher order terms, we plot $m_\pi^2/(m_x+m_y)$ in \figref{m-and-f} (left). 
As the lattice spacing decreases,
the PQ log at small mass becomes more evident.
At larger lattice spacing, the
PQ log is largely washed out by staggered taste violations.
The continuum dashed red line has sea and valence masses equal, so no PQ log is expected.  

In \figref{m-and-f} (right), we show the behavior of the decay constant. 
Extrapolating to the
continuum, setting $m'_s=m_s$,
and setting light valence and sea masses equal gives the dashed red line.  
The final result for $f_\pi$ after extrapolating $m_x,m_y\!\to\!\hat m$ is marked by 
a $+$.
The experimental result is indicated by the \circle; it comes from
the $\pi^+\to\mu^++\nu_\mu$ decay width  and
$V_{ud}=0.97377(27)$ \cite{PDG}.

All points and fit lines above have been corrected for finite volume effects using the
\rschpt\ forms at one loop.
However, it is known \cite{COLANGELO}
that finite volume effects coming from  higher orders in \chpt\ can be a large
($\sim\!50\%$) correction to the one-loop effects in the current ranges of quark mass and
volume.
We therefore study this issue directly
by comparing results on the 
spatial size $20^3$ and $28^3$
$a\!\approx\! 0.12\;$fm ensembles with
$a\hat m'=0.01$, $am_s'=0.05$.  These lattices have spatial 
length $2.4\;$fm and $3.4\;$fm, respectively. The comparison is shown in \tabref{finiteV}.

\begin{table}[t!]
%\begin{table}[b]
%\vspace{0.10in}
\begin{center}
\setlength{\tabcolsep}{5mm}
\begin{tabular}{|c|c|c|c|}
\hline
quantity & \% difference &  boosted \% diff. & 1-loop \% diff. \\
\hline                            
$af_\pi$ & 1.4(2)\% & 1.6(2)\% & 1.1\% \\
\hline
$af_K$ & 0.4(3) \% & 0.4(3) \% & 0.3\% \\
\hline                            
$(am_\pi)^2$ & -1.0(4)\% & -1.2(4)\% & -0.9\% \\
\hline
$(am_K)^2$ & -0.4(2) \% & -0.4(2) \% & -0.2\% \\
\hline
\end{tabular}
\end{center}
\vspace{-5mm}
\caption{\label{tab:finiteV}Finite volume effects.
The second column is the \% difference between
the values on the $28^3$ and the $20^3$ lattice.  In the third column, we
``boost'' the difference, to take into account the (small) further difference
between $28^3$ and infinite volume. (We use one-loop results to make the
adjustment.) The last column shows the \% difference between $20^3$
and infinite volume as predicted by one-loop \rschpt.
``Pion'' quantities are from lattice valence masses $0.005$, 
$0.005$; ``kaon'' quantities are from lattice valence masses $0.005$, $0.04$.}
\vspace{-4mm}
\end{table}

As expected from Ref.~\cite{COLANGELO}, the true
finite volume effects are larger than those predicted at one-loop, although only
for $f_\pi$ are the relative errors small enough to make the comparison unambiguous.
We define the ``residual finite volume effect''
on the $20^3$ lattice as that effect not taken into account at one-loop, \ie the difference
between columns three and four in \tabref{finiteV}.
In practice, the $20^3$, $a\hat m'=0.01$, $am_s'=0.05$,  $a\!\approx\! 0.12\;$fm lattice is close to the 
worst case in our data set, since the volumes both at the lightest sea quark masses 
and at the finest lattice spacings are larger.

Judging by the one-loop results, we expect the
overall (and hence residual) finite volume effects closest to the chiral and
continuum limits in our data set to be about half those seen in the table.
We therefore correct our data by $1/2$ the residual
finite volume effects from \tabref{finiteV}, and take the full size of the correction as
a systematic error.
We note that the size of the error determined here is
very similar to that estimated by us previously \cite{FPI04} using Ref.~\cite{COLANGELO}.

The fact that we can get good fits to the forms predicted by \rschpt\ (and not
to those of continuum \chpt\ \cite{FPI04}) is an overall test of staggered
chiral perturbation theory, including the ``replica trick'' to represent rooting.
As a more focused test of the replica trick in \rschpt, we allow $n_r$, the number of
replicas per staggered flavor, to be a free fit parameter.
If \rschpt\ is correct, we should find $n_r=1/4$.
On the low-mass data set described above, we obtain $ n_r = 0.28(2)(3),$
where the errors are statistical and systematic (describing the variation over details
of the chiral fits).
While the ability of \rschpt\ to describe rooted staggered data 
cannot prove the correctness of the rooting trick itself, it does indicate that
no problems occur in the chiral sector of the rooted theory \cite{SCHPT-OTHER}.
This is because \rschpt\ reproduces
continuum  \chpt\ in the limit $a\to 0$.

Using $r_1=0.318(7)\;$fm from $\Upsilon$ splittings, we obtain (still preliminary)
\begin{eqnarray*}\label{eq:results}
%FPI04 results:
%f_\pi & = &  129.5 \pm 0.9\pm 3.6 \;  \\
%f_K & = &  156.6 \pm 1.0\pm 3.8 \; \\
%f_K/f_\pi  & = & 1.210(4)(13)\ ,
% Lat05:
%f_\pi & = &  128.6 \pm 0.6\pm 2.5 \;  \\
%f_K & = &  154.3 \pm 0.6\pm 3.0 \; \\
%f_K/f_\pi  & = & 1.200(4)({}^{+17}_{-\phantom{1}5})\ ,
%Update 9/5/05 (drop no0.0031sea fits, include scale error)
f_\pi & = &  128.3 \pm 0.5\; {}^{+2.4}_{-3.5} \; \MeV \\
f_K & = &  154.3 \pm 0.4\; {}^{+2.1}_{-3.4}  \; \MeV \\
f_K/f_\pi  & = &  1.202(3)({}^{+\phantom{1}8}_{-14})\ ,
\end{eqnarray*}
where the errors are from statistics and lattice systematics. 
These results are consistent with our previous results \cite{FPI04}, with 20--30\% smaller errrors.
Our value for $f_\pi$ is consistent with the experimental result,
$f_\pi^{\rm expt} = 130.7\pm 0.1\pm 0.36 \; \MeV$ \cite{PDG}.

Instead of setting the scale from $\Upsilon$ splittings, we can
set the scale from $f_\pi$ itself, which gives smaller errors
for $\pi$--$K$ quantities.  Note that even dimensionless
quantities can change with the new scale, due to changes in physical quark masses.
We then obtain (preliminary):
\vspace{-6mm}
 \begin{eqnarray*}
f_K  =    156.5 \pm 0.4\; {}^{+1.0}_{-2.7}  \; \MeV  &\quad\qquad& f_K/f_\pi   = 1.197(3)({}^{+\phantom{1}6}_{-13})  \\
   f_\pi/f_{2} =  1.052(2)({}^{+6}_{-3}) &\quad\qquad& 
			\langle\bar uu\rangle_{2} =   -(\, 278(1)({}^{+2}_{-3})(5)\;\MeV\,)^3 \\
 f_\pi/f_3 = 1.21(5)({}^{+13}_{-\phantom{1}3}) &\quad\qquad& 
			\langle\bar uu\rangle_{3}  =  -(\, 242(9)({}^{+\phantom{2}5}_{-17})(4)\;\MeV\,)^3  \\
f_{2}/f_{3}  =  1.15(5)({}^{+13}_{-\phantom{1}3})  &\quad\qquad& \langle\bar uu\rangle_{2}/\langle\bar uu\rangle_{3}  =  1.52(17)({}^{+38}_{-15}) \\
2L_6 - L_4 = 0.4(1)({}^{+2}_{-3}) &\quad\qquad& 2L_8 - L_5 = -0.1(1)(1) \\
L_4 = 0.4(3)({}^{+3}_{-1}) &\quad\qquad& L_5 =2.2(2)({}^{+2}_{-1})  \\
L_6 = 0.4(2)({}^{+2}_{-1}) &\quad\qquad& L_8 =1.0(1)(1)  \\
m_s = 88(0)(3)(4)(0)\;\MeV &\quad\qquad&  \hat m = 3.2(0)(1)(2)(0)\;\MeV \\
m_u = 1.9(0)(1)(1)(1)\;\MeV &\quad\qquad&  m_d = 4.6(0)(2)(2)(1)\;\MeV \\
m_s/\hat m = 27.2(1)(3)(0)(0) &\quad\qquad&  m_u/m_d = 0.42(0)(1)(0)(4)\;.
\end{eqnarray*}
The errors are statistical, lattice-systematic, perturbative (for masses and condensates; from two-loop perturbation theory \cite{Mason:2005bj}) and
electromagnetic (for masses; from continuum estimates). 
$f_2$ ($f_3$) represents the three-flavor decay constant
in the two (three) flavor chiral limit, and
$\langle\bar uu\rangle_{2}$ ($\langle\bar uu\rangle_{3}$)
is the corresponding condensate.
The low energy constants $L_i$ are in units of $10^{-3}$
and are evaluated at chiral scale $m_\eta$; the condensates and masses are in the
$\msbar$ scheme at scale $2\,\GeV$. 

We also obtain 
\vspace{-2mm}
$$\eqn{r1}
 r_1 =0.3108(15)({}^{+26}_{-79})\; {\rm fm}\ , 
\vspace{-1mm}
$$
which is 1-$\sigma$ lower (and with somewhat smaller errors) than the value from the $\Upsilon$ system.
There is a 2-$\sigma$ conflict between our $r_1$ result from $f_\pi$
and the HPQCD Collaboration \cite{HPQCD-UPSILON} value from $\Upsilon$ splittings, 
$r_1=0.321(5)\;$fm. 
If instead we compare to our own evaluation of $r_1$ from the $\Upsilon$ spectrum,
$r_1=0.318(7)$ \cite{FPILAT05}, the difference is only $1$-$\sigma$. We emphasize, however, that the
evaluations of $r_1$ from the $\Upsilon$ splittings both by us and by the HPQCD Collaboration
use the same lattice data: HPQCD $\Upsilon$ splittings \cite{HPQCD-UPSILON}
and MILC values of $r_1/a$ \cite{MILC-SPECTRUM}. The difference is only in how
we extrapolate to the physical point and estimate the systematic error.
Our result
is consistent, though, with the ($N_f=2$) result from the {ETM Collaboration} \cite{TMfpi},
$r_0=0.454(7)\;$fm.  Converting from $r_0$ to $r_1$ using the 
ratio $r_0/r_1 = 1.46(1)(2)$ (from Ref.~\cite{MILC-SPECTRUM}, adjusted for the slight
difference between $N_f=3$ and $N_f=2$), this gives
$r_1= 0.311(7)\;$fm.

Together with the experimental result for the kaon leptonic branching fraction \cite{KLOE}, 
our result for $f_K/f_\pi$ implies
$|V_{us}|=0.2246({}^{+25}_{-13})$, which is consistent with (and competitive with) 
the world-average value $|V_{us}|=0.2257(21)$ \cite{PDG} coming from semileptonic $K$-decay
coupled with non-lattice theory.

The change in the perturbative mass renormalization
constant $Z_m$ from one to two loops accounts for almost all of the difference between the mass
values quoted here and those in Ref.~\cite{STRANGE-MASS,FPI04}. A non-perturbative evaluation
of $Z_m$ is in progress.

We stress that our extraction of the $L_i$ uses fits
that include (analytic) NNLO terms.  Therefore, a comparison to other evaluations, 
either phenomenological or on
the lattice, that stop at NLO terms is problematic.
Indeed, NNLO terms of ``natural size'' in \chpt\ can produce
changes in the $L_i$ (relative to a pure NLO evaluation) that are as large as, or even somewhat larger than, our current systematic errors.
This is confirmed by NLO fits to our data.  Such fits
have very poor confidence levels, however, which is why we do not include them in the analysis.

The $SU(2)_L\times
SU(2)_R$ LECs $\bar l_3$,
$\bar l_4$  that are extracted \cite{Leutwyler:2007ae} from our $SU(3)_L\times
SU(3)_R$ results using one-loop (NLO) formulae are therefore quite
sensitive to the NNLO terms,
particularly for $\bar l_3$.
An NLO fit, on the other hand, 
gives $\bar l_3=2.85(7)$ (statistical errors only),
which is comparable to the results from  groups \cite{TMfpi,DelDebbio} performing
two-flavor simulations with NLO $SU(2)_L\times SU(2)_R$ fits.
Indeed, this must be true, because the $m_\pi^2$ 
data are so linear (see \figref{masses}), which requires $\bar l_3$ to have
roughly this value \cite{Leutwyler:2007ae}.
Alternative fits using $SU(2)_L\times SU(2)_R$ \rschpt\ are in progress.
Since the strange sea-quark is omitted from the chiral theory,
the approach should make possible good NLO fits on light-mass data,
and thereby bypass this issue.
Inclusion of two-loop (continuum) chiral logs \cite{Bijnens:2006jv} 
in $SU(3)_L\times SU(3)_R$ fits is also in progress.

This work is supported in part by the US DoE and NSF.
Computations were performed at the NSF Teragrid, NERSC, and USQCD centers, and at computer centers
at the University of Arizona, the University
of California at Santa Barbara, Indiana University, and the University of Utah.


\vspace{-2mm}

\begin{thebibliography}{99}
\vspace{-2mm}

\bibitem{ASQTAD}
%T.~Blum \et\  [{MILC}], Phys. Rev. D {\bf 55}, 1133 (1997);
%K.~Orginos and D.~Toussaint, Phys. Rev. D {\bf 59} (1999) 014501;
% and
%Nucl. Phys. B (Proc. Suppl.) {\bf 73}, 909 (1999);
K.~Orginos, D.~Toussaint and R.L.~Sugar,
Phys. Rev. D {\bf 60} (1999) 054503 and references therein.
%and %hep-lat/9909087, %Nucl. Phys. B (Proc. Suppl.) {\bf 83-84}, 878 (2000);
%G.P.~Lepage, Nucl. Phys. (Proc. Suppl.) {\bf 60A}, 267 (1998) and
%G.P.~Lepage, Phys. Rev. D {\bf 59} (1999) 074502;
%J.F.~Lag\"ae and D.K.~Sinclair, Nucl. Phys. (Proc. Suppl.) {\bf 63}, 892 (1998);
%C.~Bernard \et\  [{MILC}], Phys. Rev. D {\bf 58} (1998) 014503.
%Phys. Rev. D {\bf 59} (1999) 014511, and
%Phys. Rev. D {\bf 61}, 111502 (2000).

\bibitem{FPI04}
C.~Aubin \et\ [MILC Collaboration],
%``Light pseudoscalar decay constants, quark masses, and low energy
  %constants from three-flavor lattice QCD,''
\prd{70}, 114501 (2004).
%[hep-lat/0407028].

\bibitem{FPILAT04}
C.~Aubin \et\ [MILC Collaboration],
\fermilabtwo\ 231. %[hep-lat/0409041].

\bibitem{FPILAT05}
C.\ Bernard \et\ [MILC Collaboration],
%``Update on pi and K Physics,''
\dublin{025},  hep-lat/0509137;
%\bibitem{FPILAT06}
%C.\ Bernard \et\ [MILC Collaboration],
%``Update on the physics of light pseudoscalar mesons,''
\tucson{163}, hep-lat/0609053; and
%\bibitem{FPIChiDyn06}
%C.\ Bernard \et\ [MILC Collaboration],
hep-lat/0611024,
to be published in the
proceedings of {\it Chiral Dynamics 2006}, Duke University, Sept. 18-22, 2006.






%\bibitem{HL-SCHPT}
%C.\ Aubin and C.\ Bernard,
%``Staggered chiral perturbation theory with heavy-light mesons,''
%\fermilabtwo\ 491, %[hep-lat/0409027], 
%and in preparation.

%\bibitem{FERMI-HL}
%C.\ Aubin \et\ [{Fermilab Lattice}, 
%MILC Collaboration, and 
%{HPQCD} Collaborations] 
%``Semileptonic decays of $D$ mesons in three-flavor lattice QCD,''
 %\prl{94}, 011601 (2005) %[hep-ph/0408306] 
%and 
%``Charmed meson decay constants
%in three flavor
%lattice QCD,''
%Phys.\ Rev.\ Lett.\ {\bf 95}, 122002 (2005)
%[hep-lat/0506030];
%E.~Gulez \et\ [{HPQCD} Collaboration],
%A.~Gray, M.~Wingate, C.~T.~H.~Davies, G.~P.~Lepage and J.~Shigemitsu,
  %``B meson semileptonic form factors from unquenched lattice QCD,''
  %Phys.\ Rev.\ D {\bf 73}, 074502 (2006)
  %[arXiv:hep-lat/0601021];
%J.\ Simone [for {Fermilab Lattice} and
%MILC Collaboration
%Collaborations], poster at this conference.

%%%%%%%%%%%%%%%%%%%%%%%%%%%%%%%%%%%%%%%%%%%%%%%


%\bibitem{SYM-GAUGE}
%K.~Symanzik, in ``Recent Developments in Gauge Theories'', eds.
%G. 't Hooft \et, 313 (Plenum, New York, 1980);
%Nucl. Phys. {\bf B226} 187 (1983);
%M.~L\"uscher and P.~Weisz, Comm. Math. Phys. {\bf 97} 19 (1985);
%Phys. Lett. {\bf 158B} 250 (1985);
%M.~Alford, W.~Dimm, G.P.~Lepage, G.~Hockney and P.B.~Mackenzie,
%Phys. Lett. {\bf 361B} 87 (1995).

\bibitem{MILC-POTENTIAL}
C.~Bernard \et\ [MILC Collaboration],
%hep-lat/0002028,
Phys. Rev. D {\bf 62}, 034503 (2000).

\bibitem{SOMMER}
R. Sommer, Nucl. Phys. {\bf B411}, 839 (1994).

\bibitem{HPQCD-UPSILON}
A.~Gray \et\  [{HPQCD Collaboration}],
%I.~Allison, C.~T.~H.~Davies, E.~Dalgic, G.~P.~Lepage, J.~Shigemitsu and M.~Wingate,
  %``The Upsilon spectrum and m(b) from full lattice QCD,''
  Phys.\ Rev.\  D {\bf 72}, 094507 (2005).
  %[arXiv:hep-lat/0507013].

\bibitem{MILC-SPECTRUM}
C.~Aubin \et\ [MILC Collaboration],
Phys.\ Rev.\ D {\bf 70}, 094505 (2004).
%[arXiv:hep-lat/0402030].

\bibitem{Sommer:2003ne}
R.~Sommer {\it et al.}  [ALPHA Collaboration],
%``Large cutoff effects of dynamical Wilson fermions,''
Nucl.\ Phys.\ Proc.\ Suppl.\  {\bf 129}, 405 (2004).
%[arXiv:hep-lat/0309171].

\bibitem{LEE-SHARPE}
W.\ Lee and S.\ Sharpe
\prd{60}, 114503 (1999);
C.\ Bernard,
Phys.\ Rev. D {\bf 65}, 054031 (2002).
%[arXiv:hep-lat/0111051].

\bibitem{SCHPT-CACB}
C.\ Aubin and C.\ Bernard, Phys.\ Rev. D {68}, 034014 (2003)
%[arXiv:hep-lat/0304014] 
and Phys.\ Rev. D {\bf 68}, 074011 (2003).
% [arXiv:hep-lat/0306026].

\bibitem{SCHPT-OTHER}
C.~Bernard,
%``Staggered Chiral Perturbation Theory and the Fourth-Root Trick,''
Phys.\ Rev.\ D {\bf 73}, 114503 (2006); % [hep-lat/0603011];
C.~Bernard, M.\ Golterman, and Y.\ Shamir, 
%``Effective field theories for rooted staggered fermions,''
\regensburg{263},
arXiv:0709.2180[hep-lat] and in preparation.

\bibitem{PDG}
W.-M. Yao  {\it et al.}\ [Particle Data Group],
Journal of Physics G {\bf 33}, 1 (2006)
and 2007 partial update.

\bibitem{COLANGELO}
%\bibitem{Colangelo:2005gd}
  G.~Colangelo, S.~D\"urr and C.~Haefeli,
  %``Finite volume effects for meson masses and decay constants,''
  Nucl.\ Phys.\  B {\bf 721}, 136 (2005).
  %[arXiv:hep-lat/0503014].

\bibitem{Mason:2005bj}
Q.~Mason \et\  [{HPQCD Collaboration}],
%H.~D. Trottier, R.~Horgan, C.~T.~H. Davies and G.~P. Lepage, {\em
Phys.\ Rev.\ D {\bf 73}, 114501 (2006).% [arXiv:hep-ph/0511160].

%\bibitem{DONOGHUE}
%We thank J.\ Donoghue for the estimate, based on the continuum
%literature.


%\bibitem{MILC-POTENTIAL-AND-SPECTRUM}
%C.~Bernard \et\ [MILC Collaboration],
%hep-lat/0002028,
%Phys. Rev. D {\bf 62}, 034503 (2000) and
%%C.~Bernard \et\ [MILC Collaboration],
%Phys. Rev. D {\bf 64} (2001) 054506; 
%% and \boston 257. %[arXiv:hep-lat/0208041].



%\bibitem{HISQfpi}
%\bibitem{Follana:2007uv}
  %E.~Follana \et\ [{HPQCD Collaboration}],
%C.~T.~H.~Davies, G.~P.~Lepage and J.~Shigemitsu  
  %``High Precision determination of the pi, K, D and D_s decay constants   from
  %lattice QCD,''
  %arXiv:0706.1726 [hep-lat].

\bibitem{TMfpi}
%bibitem{Boucaud:2007uk}
Ph.~Boucaud {\it et al.}  [ETM Collaboration],
  %``Dynamical twisted mass fermions with light quarks,''
  Phys.\ Lett.\  B {\bf 650}, 304 (2007).
  %[arXiv:hep-lat/0701012].

\bibitem{KLOE}
F.~Ambrosino {\it et al.}\  [{KLOE Collaboration}],
  %``Measurement of the absolute branching ratio for the K+ --> mu+ nu  (gamma)
  %decay with the KLOE detector,''
  Phys.\ Lett.\  B {\bf 632}, 76 (2006).
  %[arXiv:hep-ex/0509045].


\bibitem{STRANGE-MASS} C.\ Aubin \et\ [{HPQCD, MILC, and UKQCD Collaborations}],
%``First determination of the strange and light quark masses from full
  %lattice QCD,'' 
Phys.\ Rev.\ D {\bf 70}, 031504(R) (2004).
%[hep-lat/0405022].

\bibitem{Leutwyler:2007ae}
  H.~Leutwyler,
  %``Insights and puzzles in light quark physics,''
  arXiv:0706.3138 [hep-ph].

\bibitem{DelDebbio}
%\bibitem{Del Debbio:2006cn}
  L.~Del Debbio \et,
  %L.~Del Debbio, L.~Giusti, M.~Luscher, R.~Petronzio and N.~Tantalo,
  %``QCD with light Wilson quarks on fine lattices. I: First experiences and
  %physics results,''
  JHEP {\bf 0702}, 056 (2007) and
  %[arXiv:hep-lat/0610059] 
%bibitem{Del Debbio:2007pz}
  %L.~Del Debbio, L.~Giusti, M.~Luscher, R.~Petronzio and N.~Tantalo,
  %``QCD with light Wilson quarks on fine lattices. II: DD-HMC simulations and
  %data analysis,''
  JHEP {\bf 0702}, 082 (2007).
  %[arXiv:hep-lat/0701009].

\bibitem{Bijnens:2006jv}
  J.~Bijnens, N.~Danielsson and T.~A.~Lahde,
  %``Three-flavor partially quenched chiral perturbation theory at NNLO for
  %meson masses and decay constants,''
  Phys.\ Rev.\  D {\bf 73}, 074509 (2006).
  %[arXiv:hep-lat/0602003].

%\bibitem{PRL}
%C.\ Davies \et\ [{Fermilab, HPQCD, MILC, UKQCD}], Phys.\ Rev.\ Lett.\ {\bf 92} 022001 (2004).
%[arXiv:hep-lat/0304004].

%\bibitem{GASSER-LEUTWYLER}
%J.~Gasser and H.~Leutwyler, Nucl. Phys. {\bf B250}, 465 (1985).


%\bibitem{DONOGHUE}
%We thank J.\ Donoghue for this estimate, based on:
%J.\ Bijnens and J.\ Prades,
%``Electromagnetic corrections for pions and kaons: Masses and
%polarizabilities,''
%Nucl.\ Phys.\ B {\bf 490} (1997) 239;
%[arXiv:hep-ph/9610360].
%%CITATION = HEP-PH 9610360
%J.\ Donoghue and A.\ Perez,
%``The electromagnetic mass differences of pions and kaons,''
%Phys.\ Rev.\ D {\bf 55} (1997) 7075;
%[arXiv:hep-ph/9611331].
%%CITATION = HEP-PH 9611331;%%
%\cite{Donoghue:1996zn}
%\cite{Moussallam:1997xx}
%\bibitem{Moussallam:1997xx}
%B.\ Moussallam,
%``A sum rule approach to the violation of Dashen's theorem,''
%Nucl.\ Phys.\ B {\bf 504} (1997) 381.
%[arXiv:hep-ph/9701400].
%%CITATION = HEP-PH 9701400;%%




%\bibitem{QCHPT}
%S.R.\ Sharpe,
%\prd{46}, 3146 (1992);
%C.\ Bernard and M. Golterman,
%\prd{46}, 853 (1992).

%\bibitem{QUENCHDELTA}
%\bibitem{abb00}
%S. Aoki \et\ [{CP-PACS}],
%Phys. Rev. Lett. {\bf 84}, 238 (2000) and
%S.~Aoki {\it et al.}  [CP-PACS Collaboration],
  %``Light hadron spectrum and quark masses from quenched lattice QCD,''
  %Phys.\ Rev.\ D {\bf 67}, 034503 (2003);
  %[arXiv:hep-lat/0206009].
  %%CITATION = HEP-LAT 0206009;%%
%\bibitem{bde00}
%W. Bardeen \et, %A. Duncan, E. Eichten, and H. Thacker,
%Phys. Rev. {\bf D62}, 114505 (2000);
%%%\bibitem{qcdsf00}
%M. G\"{o}ckeler \et\ [{QCDSF}],
%Nucl. Phys. (Proc. Suppl.) {\bf B83}, 203 (2000);
%[hep-lat/9909160].
%\bibitem{ko00}
%S. Kim and S. Ohta,
%Phys. Rev. {\bf D61}, 074506 (2000);
%\bibitem{aok01}
%\bibitem{Aoki:2002vt}
 %Y.~Aoki \et\ [{RBC}],
%``Domain wall fermions with improved gauge actions,''
  %Phys.\ Rev.\ D {\bf 69}, 074504 (2004).
  %[arXiv:hep-lat/0211023].
  %%CITATION = HEP-LAT 0211023;%%




%\bibitem{KENTUCKY}
%\bibitem{Chen:2003im}
  %Y.~Chen {\it et al.},
  %``Chiral logarithms in quenched QCD,''
  %Phys.\ Rev.\ D {\bf 70}, 034502 (2004).
  %[arXiv:hep-lat/0304005].
  %%CITATION = HEP-LAT 0304005;%%

%\bibitem{IN-BETWEEN}
% lattice 02: hep-lat/0209099;
%C. Gattringer,
% lattice 02: hep-lat/0208056;
%\bibitem{Gattringer:2003qx}
%C. Gattringer \et\ [{RGB}],
  %``Quenched spectroscopy with fixed-point and chirally improved fermions,''
  %Nucl.\ Phys.\ B {\bf 677}, 3 (2004);
  %[arXiv:hep-lat/0307013].
  %%CITATION = HEP-LAT 0307013;%%
%PHD thesis: S. Hauswirth,
%hep-lat/0204015;
%T.W. Chiu and T.H. Hsieh,
%Phys. Rev. {\bf D66}, 014506 (2002).

%\bibitem{MASON-ALPHA}
%Q.\ Mason \et\ [{HPQCD and UKQCD}], hep-lat/0503005 
%and \fermilabtwo, 713;
%C.~Davies \et\ [{HPQCD and UKQCD}],
%%``The determination of alpha(s) from lattice QCD with 2+1 flavors of dynamical quarks,''
%Nucl.\ Phys.\ Proc.\ Suppl.\  {\bf 119}, 595 (2003).
%%[arXiv:hep-lat/0209122].

%\bibitem{MUZERO}
%D.~B.~Kaplan and A.~V.~Manohar,
%Phys.\ Rev.\ Lett.\  {\bf 56}, 2004 (1986);
%A.~G.~Cohen, D.~B.~Kaplan and A.~E.~Nelson,
%JHEP {\bf 9911}, 027 (1999).
%%[arXiv:hep-lat/9909091].

%\bibitem{STRONG-CP}
%G.\ 't Hooft, Phys.\ Rev.\ Lett.\ {\bf 37}, 
%8 (1976), and  Phys.\ Rev.\ D {\bf 14},
%3432 (1976); R.\ Jackiw and C.\ Rebbi,   
%Phys.\ Rev.\ Lett.\ {\bf 37}, 172 (1976);
%C.\ Callan, R.\ Dashen, and D.\ Gross, Phys.\ Lett.\ {\bf 63B}, 334 (1976).

%\bibitem{LAT-EM}
%\bibitem{Duncan:1996xy}
  %A.~Duncan, E.~Eichten and H.~Thacker,
  %%``Electromagnetic Splittings and Light Quark Masses in Lattice QCD,''
  %Phys.\ Rev.\ Lett.\  {\bf 76}, 3894 (1996).
  %[arXiv:hep-lat/9602005].
  %%CITATION = HEP-LAT 9602005;%%

%\bibitem{AXION}
%F.\ Wilczek, \prl{40}, 279 (1978); R.\ Peccei and H.\ Quinn, \prl{38}, 1440
%(1978); S.\ Weinberg, \prl{40}, 223 (1978); T.\ Goldman and C.\ Hoffman,
%\prl{40}, 220 (1978).
%
%\bibitem{FOURTH-ROOT}
%\bibitem{Durr:2003xs}
%S.~Durr and C.~Hoelbling,
%``Staggered versus overlap fermions: A study in the Schwinger model with  N(f)
%= 0,1,2,''
%Phys.\ Rev.\ D {\bf 69}, 034503 (2004);
% [arXiv:hep-lat/0311002];
%
%\bibitem{Follana:2004sz}
%E.~Follana, A.~Hart and C.~T.~H.~Davies  [{HPQCD}],
%``The index theorem and universality properties of the low-lying eigenvalues of%improved staggered quarks,''
%Phys.\ Rev.\ Lett.\  {\bf 93}, 241601 (2004);
%[arXiv:hep-lat/0406010];
%
%\bibitem{Durr:2004as}
%S.~Durr, C.~Hoelbling and U.~Wenger,
%``Staggered eigenvalue mimicry,''
%Phys.\ Rev.\ D {\bf 70}, 094502 (2004);
%[arXiv:hep-lat/0406027];
%
%\bibitem{Adams:2003rm}
%D.~H.~Adams,
%``A simplified test of universality in lattice QCD,''
%Phys.\ Rev.\ Lett.\  {\bf 92}, 162002 (2004)
%[arXiv:hep-lat/0312025]
%CB 6/6
%and hep-lat/0411030;
%CB 6/6
%F.~Maresca and M.~Peardon,
%hep-lat/0411029;
%
%\cite{Shamir:2004zc}
%\bibitem{Shamir:2004zc}
%Y.~Shamir,
%``Locality of the fourth root of the staggered-fermion determinant:
%Renormalization-group approach,''
%Phys.\ Rev.\ D {\bf 71}, 034509 (2005).
%[arXiv:hep-lat/0412014].



%\bibitem{ZM1}
%J.~Hein \et\ [HPQCD]
%%J.~Hein, Q.~Mason, G.~P.~Lepage and H.~Trottier,
%%``Mass renormalisation for improved staggered quarks,''
%Nucl.\ Phys.\ Proc.\ Suppl.\  {\bf 106}, 236 (2002),
%[arXiv:hep-lat/0110045] and
%%J.~Hein, C.~Davies, G.~P.~Lepage, Q.~Mason and H.~Trottier  [HPQCD
%%                  Collaboration],
%%``On the strange quark mass with improved staggered quarks,''
%Nucl.\ Phys.\ Proc.\ Suppl.\  {\bf 119}, 317 (2003)
  %[arXiv:hep-lat/0209077];


%\bibitem{ZM2}
%\bibitem{Becher:2003fu}
  %T.~Becher and K.~Melnikov,
  %``The self-energy of improved staggered quarks,''
  %Phys.\ Rev.\  D {\bf 68}, 014506 (2003)
  %[arXiv:hep-lat/0302014].

%\bibitem{SUMRULE}
%See, for instance, R.~Gupta and K.~Maltman,
%%``Light quark masses: A status report at DPF 2000,''
%Int.\ J.\ Mod.\ Phys.\ A {\bf 16S1B}, 591 (2001).
%%[arXiv:hep-ph/0101132].



      \end{thebibliography}
\end{document}